# Usability Investigation on the Localization of Text CAPTCHAs: Take Chinese Characters as a Case Study

Junnan Yu*, Xuna Ma, Ting Han**

School of Media & Design, Shanghai Jiao Tong University, Shanghai, China

*Junius@sjtu.edu.cn, **Hanting@sjtu.edu.cn

**Abstract**

Text CAPTCHA has been an effective means to protect online systems from spams and abuses caused by automatic scripts which pretend to be human beings. However, nearly all the Text CAPTCHA designs in nowadays are based on English characters, which may not be the most user-friendly option for non-English speakers. Therefore, under the background of globalization, there is an increasing interest in designing local-language CAPTCHA, which is expected to be more usable for native speakers. However, systematic studies on the usability of localized CAPTCHAs are rare, and a general procedure for the design of usable localized CAPTCHA is still unavailable. Here, we comprehensively explored the design of CAPTCHAs based on Chinese characters from a usability perspective: cognitive processes of solving alphanumeric and Chinese CAPTCHAs are analyzed, followed by a usability comparison of those two types of CAPTCHAs and the evaluation of intrinsic design factors of Chinese CAPTCHAs. It was found that Chinese CAPTCHAs could be equally usable comparing with alphanumeric ones. Meanwhile, guidelines for the design of usable Chinese CAPTCHAs were also presented. Moreover, those design practices were also summarized as a general procedure which is expected to be applicable for the design of CAPTCHAs based on other languages.

## I. Introduction

Since its invention in 2002, CAPTCHA (Completely Automated Public Turing test to tell Computers and Humans Apart) has been an effective means to defend automatic scripts and protect online systems from spam and abuse, particularly in registration and password verification scenes [1, 2]. For instance, CAPTCHA has been deployed by Gmail and Facebook to filter out fake registrations and/or defend violent crack of user passwords. Also, PayPal and online banking systems utilize it to enforce financial security for their clients. Principally, a well-designed CAPTCHA is easy for human to recognize while hard for bots to crack. In nowadays, CAPTCHAs employed by online systems are mainly based on Image or Text [3]. For Image CAPTCHAs, a set of images are presented each time and the user is instructed to click on particular image(s) to solve them [4]. A typical Text CAPTCHA usually includes several alphanumeric characters (a-z, A-Z, 0-9) that are distorted and added with background noises to reduce its possibility of being recognized by computer algorithms [5]. Given that this paper focuses on Text CAPTCHA, the word CAPTCHA mentioned afterwards represents only the text one unless otherwise specified. During the verification process, human users are required to correctly input the characters appeared on a CAPTCHA [6]. However, with the increased distortion of texts and background, it is more efficient to defend automatic scripts [5]  but also at the cost of degraded usability. Therefore, several studies have been conducted to investigate the usability of alphanumeric CAPTCHAs: Chellapilla [7] presented the limits of distortions that are acceptable for users, which is a comprehensive large-scale evaluation of the design factors; Elie Bursztein [8] investigated the effects of visual features, anti-segmentation features, and anti-recognition features on the usability of CAPTCHAs. Lee [9] found that the usability of CAPTCHA varies with the age of users and the young group performed better than the old group. Belk [10] revealed that the participant's cognitive styles also affect the usability of a particular CAPTCHA and they suggested that users' cognitive styles and culture backgrounds should be considered when designing CAPTCHAs.

In addition to those usability studies on alphanumeric CAPTCHAs, there is also an increasing importance to design localized CAPTCHAs that use regional languages due to the large number of online users who are non-English speakers. Meanwhile, people are also intuitively comfortable with their native languages. Localized CAPTCHAs could have several advantages: (i) Familiar for local users: it is intuitively more comfortable with native languages and a better usability would be expected; (ii) More durable with design factors: for local languages that are complex in form, it may not be easily confused with security features such as distortion or background noises; (iii) More candidate characters and better security: there are 26 letters and 10 numbers for alphanumeric CAPTCHAs, however, take Chinese Language for example, it can easily provide thousands of different candidate characters, which is expected to provide better security because of the complexity of various characters. For instance, Shirali-Shahreza [11] designed a CAPTCHA mechanism that employs Persian/Arabic characters. Yang [12] explored the application of Korean characters in CAPTCHAs, and found they are easier for native Korean speakers. Banday [13] investigated the possibility of developing CAPTCHAs based on Urdu, one of the regional languages used in India. Accompanying with the deployments of Chinese characters by some leading internet companies, such as Baidu.com and Renren.com, there is also an emerging interest in developing new algorithms for the generating of Chinese CAPTCHAs. Wang [14] proposed double-layer Chinese CAPTCHAs against OCR (Optical Characters Recognition), and a couple of other methods for the generation of Chinese CAPTCHAs are also reported [15-17].

However, those previous studies are mostly focused on new means of generating localized CAPTCHAs while lack systematical investigations on the usability of such localized CAPTCHAs. Particularly, the general procedure for designing localized CAPTCHAs is still unavailable yet. In this study, taking the usability investigation of Chinese CAPTCHAs as a case study, we analyzed the cognitive processes of solving English and Chinese CAPTCHAs, compared the usability of CAPTCHAs based on English and Characters, and explored the intrinsic factors that may affect the usability of Chinese CAPTCHAs. Such comprehensive practices on Chinese CAPTCHA were further generalized as a standard procedure, which is expected to be applicable for the localization of CAPTCHAs based on any other regional languages.

The rest of this paper is organized as follows: Part II summarized related works on the design of localized CAPTCHAs; Part III and Part IV introduced the two experimental designs; Part V presented and discussed the experimental results; Part VI proposed a generalized procedure for the localization of CAPTCHAs; Part VII presented the conclusion.

## II. Related Work

Previous works mainly focused on the design of localized CAPTCHAs. For example, Shirali-Shahreza [11] designed a CAPTCHA that employed Persian/Arabic characters, which are widely used by all Muslim communities and usually complex due to the difference of letter size, lack of space between words, variation of formats, etc. Their study indicated that, integrating those inherent complexities with random noises and backgrounds, it can hardly be recognized by commercial Arabic OCR (Optical Characters Recognition) software and is therefore secure. Meanwhile, user study demonstrated a high recognition rate of 90% for human beings. Yang [12] explored the application of Korean syllables in text CAPTCHAs, in which each character was rotated, split, distorted and combined with a background image that contained random lines and special effects. They concluded that such a design is secure and effective for Korean users. Fidas [18] compared the solving time and correction rate of text CAPTCHAs using English and non-English Characters. Their study implied that, for native speakers, the correction rate of native-language CAPTCHAs is better than that of English ones, while the solving time are similar for CATPCHAs based on both languages. Banday [13] investigated the deployment of CAPTCHAs based on Urdu, one of the regional languages used in India. Their preliminary results indicated that, for native speakers of Urdu who had little or no familiarity with English, they solved Urdu CAPTCHAs significantly faster and more accurately than those based on English.

The studies of CAPTCHAs based on Chinese are mainly focused on enhancing security issues. For instance, To hack the touclick Chinese CAPTCHA, Shen [19] developed an algorithm utilizing multi-scale Corner based Structure Model, which is efficient to capture the structure of Chinese characters. Wang [14] proposed a design which incorporated a semi-transparent layer of Chinese characters as the background of main layer and experimentally proved that it is effective against OCR. Chen [15] proposed a Chinese CAPTCHA model based on AJAX. In such a design, users are required to follow the randomly generated natural language instructions and partially input the characters shown on a CAPTCHA. This design turned out to be usable while more secure.

## III. Experimental Design

The usability was evaluated by three independent variables [20]: effectiveness, efficiency and satisfaction. The effectiveness and efficiency were measured by the average solving time and correction rate for each type of CAPTCHA, respectively. The satisfaction was obtained through an online questionnaire and face-to-face interview with each participant. The first part of this section discussed the theoretical analysis: the cognition process of solving CAPTCHAs based on English and Chinese characters. The second part introduced the common features of generating

CAPTCHAs, such as distortion, overlapping, noise level, etc. The third part focused on Experiment I: comparing the usability of English and Chinese CAPTCHAs. The fourth part detailed Experiment II, which took into account the findings in Experiment I to evaluate the intrinsic factors that may affect the usability of Chinese CAPTCHAs.

## 3.1. Cognition process of solving English and Chinese CAPTCHAs

The cognition process of solving CAPTCHAs by a standard keyboard generally involves three steps: visual input, information processing and action, which is summarized in Fig. 1. For CAPTCHAs that require to input English characters, the solving process is straightforward because all letters and numbers are explicitly arranged on a standard keyboard. Therefore, after visual caption of an English CAPTCHA, one would recognize each individual character, recall the correspondent locations of keys, and finally strike the correspondent keys.

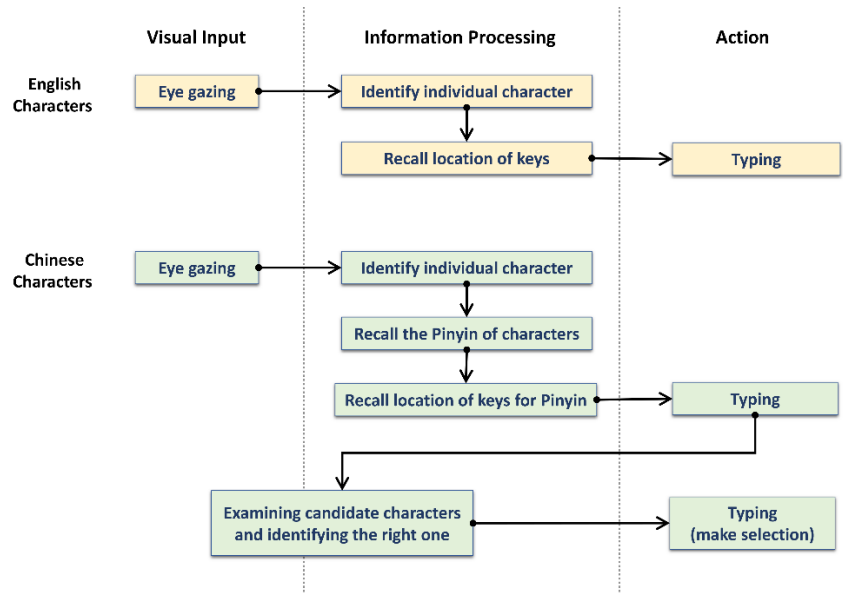

**Fig. 1**. Cognition process toward solving English and Chinese CAPTCHAs with standard keyboard.

However, the typing of Chinese characters is different from that of English because it is a non-alphabetic language. For the input of Chinese CAPTCHAs, a coding system, Pinyin, is generally used to represent tens of thousands of Chinese Characters with only 26 alphabets. The Pinyin of a character is essentially the phonetic symbol of that character and it is mastered by any native speakers. Given that each Chinese character is labeled with the combination of serval English letters based on the pronunciation of that character, it is quite often that the same alphabetic combination usually corresponds to several Chinese characters. Therefore, the Pinyin input method generally involves selecting from several candidate characters. As shown in Fig. 2, take the Pinyin "tong" for example, several candidate characters are presented for selection. Although the input of Chinese character involves more steps, it is actually very efficient for native speakers because of the intensive daily usage of Pinyin.

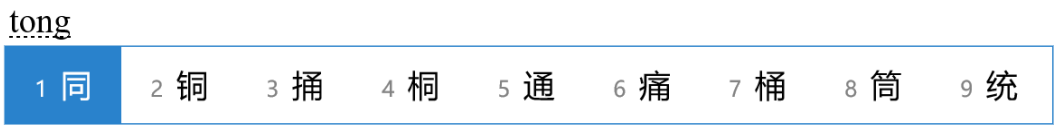

**Fig. 2**. Demonstration of Microsoft Pinyin, a popular input method of Chinese Characters and also the input method used in this study. The keyboard input of this demonstration is "tong" and nine candidate Chinese characters are listed in the frame below and labeled with numbers for making a selection. Another nine candidate characters will be presented though pressing "Page Up" or "Page Down" on a keyboard.

### 3.2. Generation of CAPTCHAs

The CAPTCHAs in this study were generated through the revision of Securimage [21], a widely-used open source code. Besides the characters used in each CAPTCHA, all other design factors were kept the same. For instance, each CAPTCHA was 230 pixels in width and 70 pixels in height. The font size was the same for all designs and the font family was Microsoft Yahei unless otherwise specified, which supports both English and Chinese characters. The transparency of characters displayed on each CAPTCHA were set as 25%. Each CAPTCHA includes 3 random lines and the background noise levels, and the distortion of each character were also kept the same for all CAPTCHAs. Furthermore, although each English CAPTCHA included 8 letters while the Chinese one included 3 or 4 Chinese characters, the averaged keystrokes [22] required for their inputs were the same under current experimental setting. Therefore, it maintained a similar workload to input different CAPTCHA types and provided a similar condition to evaluate the solving time of different designs.

### 3.3. Experiment I: usability comparison between English and Chinese CAPTCHAs.

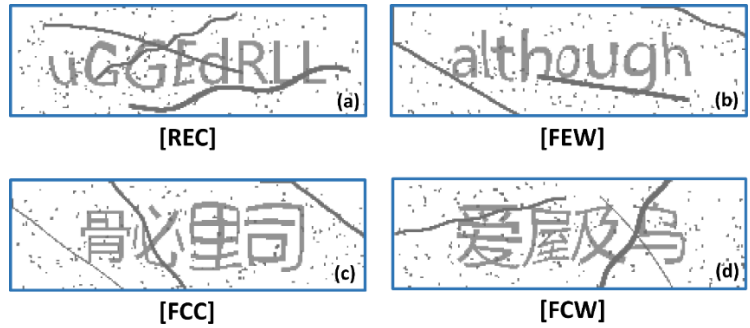

**Fig. 3.** Illustration of CAPTCHA styles explored in the usability comparison of English and Chinese in current study: (a) Random English Characters (REC); (b) Frequent English Word (FEW); (c) Random Chinese Characters (RCC); (d) Frequent Chinese Word (FCW).

The usability comparison was performed under a similar design frame, in which the only difference is the random characters or words employed on each CAPTCHA. As shown in Fig. 3, four kinds of CAPTCHAs were deployed to compare the usability of English and Chinese CAPTCHAs: Random English Characters (REC), Frequent English Words (FEW), Random Chinese Characters (RCC) and Frequent Chinese Words (FCW). Each REC CAPTCHA includes 8 English letters and each FEC CAPTCHA includes a single English word that is consisted 8 letters. The RCC CAPTCHA is the counterpart of REC CAPTCHA and it randomly includes 3 or 4 individual Chinese characters while the FCW CAPTCHA is the counterpart of FEW CAPTCHA and it includes a single Chinese word that is composed of 4 individual Chinese characters. For each language, the characters or words used were those frequently used in daily life.

### 3.4. Experiment II: Evaluation of intrinsic design factors that may affect the usability of Chinese CAPTCHAs

This experiment focused on the unique design factors of Chinese characters, which includes different font families, characters that are similar in form or pronunciation, characters of different usage frequency, etc. The four fonts studied here were shown in Fig. 4, which are (a) Yahei, (b) Songti, (c) Heiti and (d) Caoshu. The first three fonts are widely used in daily lives and have better readability comparing with the last one, Caoshu. Among all those fonts, Songti is the slimmest one while Heiti is the boldest one. The slimness and boldness of Yahei are in the middle of Songti and Heiti but less than Caoshu.

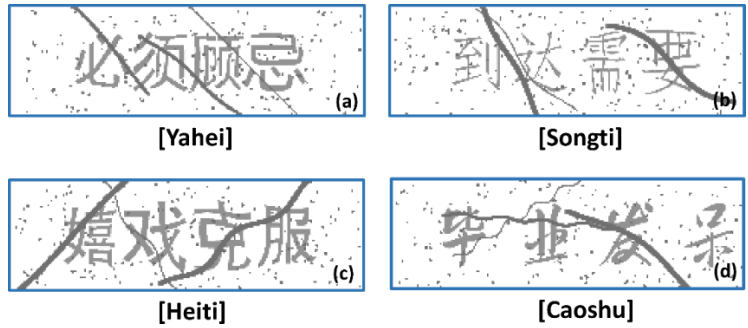

**Fig. 4**. Illustration of Chinese CAPTCHA designs based on four different fonts: (a) Yahei, (b) Songti, (c) Heiti and (d) Caoshu.

In addition to the fonts, the other four design factors are as follows: CAPTCHAs based on less frequently-used characters and characters with similar appearance are shown in Fig. 5 (a) and (b), respectively. As discussed before, the input of Chinese relays on Pinyin, which is actually the alphabetic coding of a character according to its pronunciation. Generally, the Pinyin of a Chinese character includes two parts: initial consonant and simple or compound vowel. However, the pronunciation of several pairs of initial consent, "z" and "zh" for example, are similar, which may mislead the input of characters that contain those initial consents. It is the same case for simple or compound vowels. Therefore, the last two factors focused on the effect of similar pronunciation (easily-confused Pinyin): Fig. 5 (c) and Fig. 5 (d) are examples of CAPTCHAs that employ characters with similar initial consonants (z/zh, c/ch, s/sh, r/l, l/n, f/h) and similar simple or compound vowels (an/ang, en/eng, in/ing), respectively.

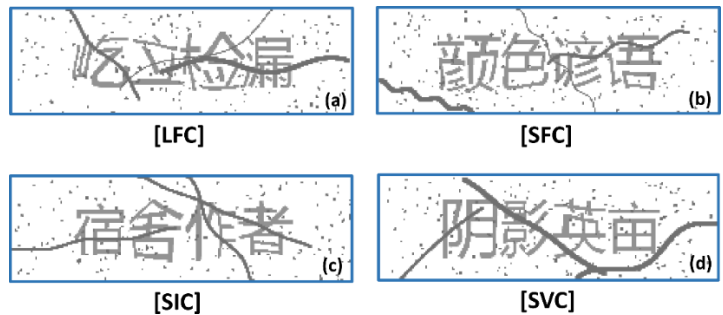

**Fig. 5**. Example of CAPTCHAs based on (a) less-frequent characters [LFC], (b) similar-form characters [SFC], (c) characters with similar initial consonants [SIC], and (d) characters with similar simple or compound vowels [SVC].

# IV. Experimental Method

## 4.1. Participants

Thirty participants (13 males and 17 females), who are native speakers of Chinese with English as a familiar second language, were recruited for Experiment I. Their age ranged from 18 to 25 (M=21.6, SD=1.3) and they were students at Shanghai Jiao Tong University (9 undergraduates and 21 graduates) students. All those participants had passed the College English Test Band 6, a language proficiency test held by the Ministry of Education of China. Therefore, they were familiar with all the English words used in Experiment I. Another 30 participants (14 males and 16 females, native speakers of Chinese) were recruited for Experiment II. Their age range from 18 to 25 (M = 21.4, SD = 2.17) and they were also students at Shanghai Jiao Tong University (16 undergraduates and 14 graduates). All participants recruited for both Experiment I and II were experienced computer users who spent at least 2 hours per week on word processing with keyboard and mouse as the primary input device. Furthermore, all subjects had encountered English CAPTCHAs and Chinese CAPTCHAs during their previous online activities before participating current experiments. Also, none of the participants had trouble reading on the screen or operating the computer input devices.

## 4.2. Apparatus

The two experiments were both conducted in a controlled lab environment. All participants were instructed to solve CAPTCHAs on a same setup, which included a 20-inch liquid crystal display with a resolution of $1440 \times 900$, a computer running Windows 8.1 system, a set of regular QWERTY keyboard and mouse as input devices. Microsoft Pinyin was used for the typing of Chinese characters. It is the pre-installed input method of Windows 8.1 and all the participants used it in daily lives. Participants were encouraged to comfort themselves by adjusting the tilt angle, height of the computer display, as well as the chair positon. All CAPTCHAs were generated on a remote server and downloaded in the form of webpages to the local browser, which was Google Chrome in this study. After solving CAPTCHAs of each session, each participant was required to finish an online questionnaire and interviewed to learn their subjective opinions regarding those CAPTCHA designs.

## 4.3. Tasks

All participants were required to finish three consecutive tasks: Firstly, each participant was instructed to solve five test CAPTCHAs to get familiar with the experimental apparatuses. Secondly, different types of CAPTCHA design are presented for participants to solve. Each type of design used 12 randomly generated CAPTCHAs for averaging. Finally, participants filled in an online questionnaire and were interviewed to learn their subjective perceptions on the different type of CAPTCHA designs.

## 4.4. Procedure

The experiments were carried out in three stages—experiment preparation, testing, and interview. During the preparation stage, the apparatuses were reset and each participant was informed with the experimental purposes and tasks. All participants were also informed that the test was anonymous and any data collected would be restricted for the use of current study only. After that, the participant was instructed to get familiar with the experiment apparatuses by means of solving five CAPTCHAs that were prepared for testing purpose. During the test session, the participant was left alone in the lab and different types of CAPTCHA design are presented one by one through the web interface. Each type of design included 12 randomly generated CAPTCHAs and participants were instructed to recognize, input and submit the characters shown on a CAPTCHA, which simulated the procedure of solving CAPTCHAs that are used by most websites in nowadays. After the submission of each CAPTCHA result, a record will be generated on the remote server, indexing the solving time, the user input values and whether the CAPTCHA was solved correctly. The webpage also refreshed automatically after submission and the participant was directed to solve the next CAPTCHA till the end of the task cycle. For each type of CAPTCHA design, 360 records were collected to obtain the average solving time and correction rate for that type of design. After solving all the CAPTCHAs, an online questionnaire was presented for the participant to fill in and participants were also interviewed to learn their comments and emotional feelings about these different CAPTCHA designs. After that, each participant was paid to appreciate his/her cooperation. During the test session, the experiment instructor waited outside the lab all the time in case any tech support was need.

# V. Results and Discussion

## 5.1. Usability Comparison of English and Chinese CAPTCHAs

The average solving time and correction rate for all four kinds of CAPTCHA design, which based on Random English Characters (REC), Frequent English Words (FEW), Random Chinese Characters (RCC), or Frequent Chinese Words (FCW), were illustrated in Fig. 6. A repeated measures ANOVA with a Greenhouse-Geisser correction indicated that the differences of solving time were statistically significant [$F_{(2.15, 822.57)}=305.44$, $p<0.001$].

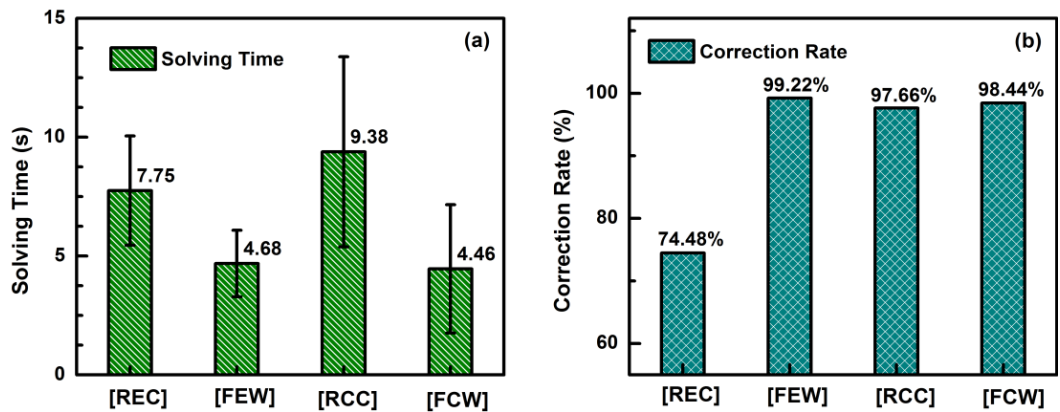

**Fig. 6.** (a) Average Solving Time and (b) Correct Rate for four kinds of CAPTCHA design: Random English Characters (REC), Frequent English Words (FEW), Random Chinese Characters (RCC), Frequent Chinese Words (FCW).

The solving time of FEW (M = 4.68s, SD = 1.4s) is essentially the same as that of the FCW (M = 4.46s, SD = 2.7s). This is attributed to the fact that all participants were familiar with both the English and Chinese words used in this study. Therefore, they responded equally fast for CAPTCHAs based on words of both languages. It is also indicated in Fig. 6 that, CAPTCHAs based on RCC (M = 9.38s, SD = 4s) have the longest solving time, followed by that of REC (M = 7.75s, SD = 2.3s). Meanwhile, the solving time of both RCC and REC are longer than that of FEC and FCC. This is because that, for both languages, it took a longer time for participants to recognize and type individually each random character. On the contrary, the characters on a FEC or FCC CAPTCHA can be recognized in whole as a meaningful word. Therefore, it took less time to solve these two kinds of CAPTCHAs. The similar solving time for both FEC and FCC further shows that it took basically the same effort for participants to response to their native language and a familiar second language. Those results reveal that CAPTCHAs based on frequently-used English and Chinese words have better efficiency than those employ random characters while there is no significant difference in the solving time of frequent English and Chinese words.

The effectiveness of those four CAPTCHA designs are represented by the percentage of CAPTCHAs that were correctly solved. As shown in Fig. 6 (b), accuracy for FEW (99.22%), RCC (97.66%) and FCW (98.44%) are similar and significantly higher than REC (74.68%). The high correct rate for the first three kinds of CAPTCHAs demonstrate that there was no difficulty for participants to correctly recognize CAPTCHAs of both languages, which is expected as they were bilingual speakers of both English and Chinese. The low correct rate of CAPTCHAs based on REC is further explored by individually analyzing the input of each participant. The results turn out that, a majority of those incorrect inputs were caused by English letters that look similar, such as "I" and "L". Furthermore, the distortion of those letters, which was aimed at an improved security to defense bots, made it even harder for users to distinguish them. When we removed the results from CAPTCHAs that contained confusing letters, the accuracy of REC was improved from 74.48% to 85.31%. However, such an accuracy is still lower than the other three CAPTCHA designs. This is further attributed to the random lines, background noises and distortions on a CAPTCHA. Although those features are helpful to improve the security of a CAPTCHA, they can partially merge or overlap with the candidate English characters and make it confusing to recognize letters individually on a REC CAPTCHA. While for FEW CAPTCHAs, even if one or two letters of a word were masked, it may still be recognized as a whole and solved correctly to maintain a high correct rate of 99.22%. Therefore, FEW design is less sensitive to the security features such as random lines or background noises, comparing with that of REC. In addition, due to the complexity of Chinese characters, even if a large portion of a Chinese character was masked by random lines or background noises, not much difficulty was caused for participants to recognize it as a whole and solve it correctly. Briefly, in terms of effectiveness, FEW, RCC and FCW CAPTCHAs are similar and much better than REC ones.

In addition to the efficiency and effectiveness studies, each participant was also instructed to finish a questionnaire and interviewed to acquire their subjective opinions toward those four types of CAPTCHA designs. The satisfactory questionnaire, which used the 5-point Likert-scale (1=strongly disagree, 5=strongly agree), focused on the following three aspects: Q1. It is visually comfortable; Q2. It's easy and efficient to recognize and input; Q3. It's appropriate for wide application. Those questionnaire results are displayed in Table 1. Friedman tests prove statistical differences in visual comfort (Q1, $\chi2(2)$=79.06, $p$<0.001), ease of use (Q2, $\chi2(2)$=77.27, $p$<0.001), and appropriateness for application (Q3, $\chi2(2)$=75.84, $p$<0.001).

**Table 1.** Satisfaction of CAPTCHAs based on Random English Characters (REC), Frequent English Words (FEW), Random Chinese Characters (RCC), Frequent Chinese Words (FCW). The questionnaire uses a 5-point Likert-scale (1=strongly disagree, 5=strongly agree).

| | REC | | FEW | | RCC | | FCW | |
|---|---|---|---|---|---|---|---|---|
| | AVG | SD | AVG | SD | AVG | SD | AVG | SD |
| **Q1.** It's visually comfortable | 3.03 | 1.13 | 4.13 | 0.78 | 2.87 | 1.17 | 4.13 | 0.78 |
| **Q2.** It's easy and efficient to recognize and input | 2.93 | 1.20 | 4.60 | 0.50 | 2.53 | 1.22 | 4.40 | 0.72 |
| **Q3.** It's appropriate for wide application | 2.63 | 1.13 | 4.50 | 0.86 | 2.07 | 1.23 | 4.20 | 0.87 |

The results indicate that FEW and FCW were the most preferred CAPTCHAs, while REC and RCC were negatively rated. The face-to-face interview further revealed that, more than 97.3% of the participants believed it was easy to recognize FEW and FCW CAPTCHAs with just a single glance. On the contrary, for CAPTCHAs based on random characters, it took more efforts to recognize each character individually. The portion of participants who were in favor of CAPTCHAs based on English or Chinese are 56.07% or 43.3%, respectively. Those who preferred English words felt it was more natural and straightforward to type English words because they did not need to switch the input method between English and Chinese. For those who preferred CAPTCHAs based on Chinese words, they felt more comfortable with native language and the Pinyin input methods in nowadays are well engineered to provide a fast input of Chinese. Although more than 78% of the participants believed that CAPTCHAs based on random Chinese characters provided the most security, there were hardly any participant who was willing to encounter such type of CAPTCHAs.

## 5.2. Intrinsic design factors that may affect the usability of Chinese CAPTCHA

Experiment I indicated that, for native speakers, the usability of Chinese CAPTCHAs could be the same as those English ones in terms of efficiency, effectiveness, and satisfactory. Therefore, we further explored the design factors that may affect the usability of those Chinese CAPTCHAs. However, we mainly focused on the intrinsic factors imposed only on Chinese characters rather than external factors such as background noises, strikethrough lines, or distortions, etc., which had been extensively studied in previous publications, like [5, 8]. Also, each Chinse CAPTCHA studied in Experiment II contained two random words and each word was consisted of two individual Chinese characters. This is a compromise between usability and security: Random Chinese characters boost the highest security but less preferred by users; A word contains four individual Chinese characters is easier for users while less secure.

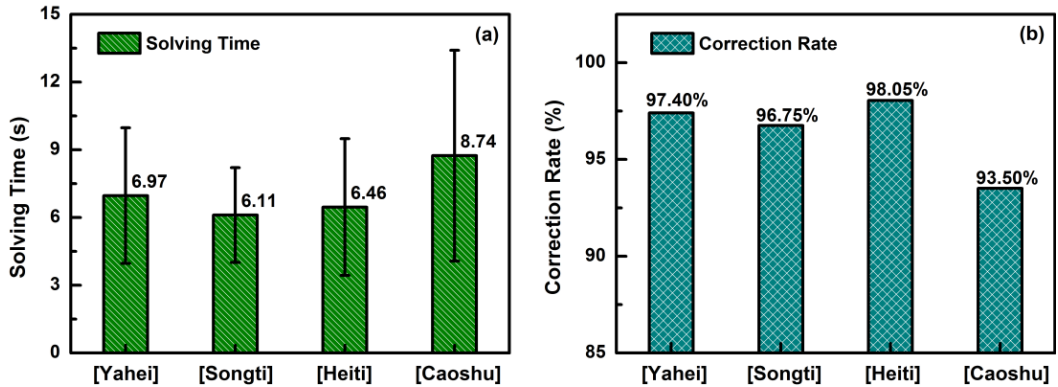

**Fig. 7**. (a) Solving time and (b) correction rate of CAPTCHAs based on four different fonts: Yahei, Songti, Heiti and Caoshu.

The solving time and correction rate for the four different fonts, Yahei, Songti, Heiti and Caoshu, were introduced in Fig. 7. A repeated measures ANOVA with Greenhouse-Geisser correction indicated that the differences of solving time are statistically significant [$F(2.45, 377.42)=20.53$, $p < 0.001$]. The solving time of Yahei (M = 6.97s, SD =3.01s), Songti (M = 6.11s, SD =2.11s), and Heiti (M = 6.46s, SD =3.04s) are similar while short than that of Caoshu (M=8.74s, SD =4.69s). The correction rate of Yahei (97.40%), Songti (96.75), and Heiti (98.05%) are also similar while higher than that of Caoshu (93.5%). Those results indicate that the efficiency and effectiveness were the same for the first three fonts and better than that of Caoshu. Because the most prominent difference between the first three fonts is the boldness of the character, this indicates that the usability of Chinese CAPTCHA is insensitive to the thickness of characters. Due to the lower readability of Caoshu, which is the counterpart of German Script fonts for English, the solving time and correction rate of Caoshu are longer and lower, respectively, comparing with the rest three fonts. The results of Yahei is chosen as the reference group because it was the default font used for the evaluation of all other Chinese CAPTCHA designs, which were shown in Fig. 6 and Fig. 8. The solving time of Yahei (M=6.97s, SD=3.01s) is between that of FCW and RCC in Fig. 6. This agrees with the fact that, to balance between usability and security, each CAPTCHA studied in Experiment II contains two random Chinese words and therefore the complexity of such CAPTCHAs is in between that of the FCW and RCC ones in Fig. 6.

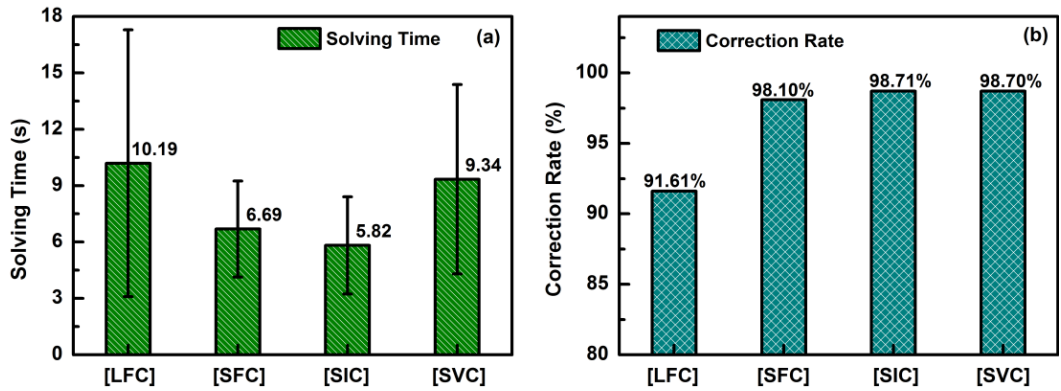

**Fig. 8**. Solving time and correct rate of Chinese CAPTCHAs based on (a) low frequency characters [LFC], (b) similar form characters [SFC], (c) similar initial consonant characters [SIC] and (d) similar vowel characters [SVC].

In addition to font, the other four factors explored in Experiment II were shown in Fig. 8. A repeated measures ANOVA indicated that the differences of the solving times among the four factors and the reference group (Yahei shown in Fig. 7) were statistically significant [$F(2.08, 317.6)=28.58$, $p<0.001$)]. For CAPTCHAs employed low-frequency characters (LFC), the solving time (M=10.19s, SD=7.77s) and correction rate (91.61%) is longer and lower comparing with that of the reference group, Yahei in Fig. 7, respectively. This reveals that, CAPTCHAs based on unfrequently used characters have lower usability in terms of efficiency and effectiveness. For CAPTCHAs based on similar-form characters (SFC), there were at least two characters looked similar. The solving time (M=6.69s, SD=2.57s) and correct rate (98.10%) of SFC have no much difference with respect to the reference group of Yahei in Fig. 7. This means that, even if the candidate characters shown on a CAPTCHAs look similar, no notable confusion or difficulties would be imposed on the solving effort. The last two factors focused on the characters that have similar pronunciation. As discussed before, the most popular means of typing Chinese is Pinyin, which is an alphabetic coding of each character according to its pronunciation. Therefore, if the pronunciation of two candidate characters are similar, their correspondent Pinyin would also be similar, which may affect the solving of a CAPTCHA. Given that the Pinyin of a character is usually composed of two parts: the initial consonants and the simple or compound vowels, the last two factors mainly focus on the CAPTCHAs that contain similar initial consonant characters [SIC] or similar vowel characters [SVC]. For SIC CAPTCHAs, the solving time (M=5.82s, SD=2.60s) and accuracy (98.71%) are similar with that of the reference group of Yahei in Fig. 7. However, for SVC CAPTCHAs, the solving time (M=9.34s, SD=5.05s) is slightly longer than the reference group of Yahei in Fig. 7. This implies that the participants had been confused to some extent and it took longer time to figure out the correct vowels. However, the accuracy remains high (98.70%), which is because the participants were familiar with the appearance of the characters and therefore could recognize the characters correctly.

During Experiment II, participants were also interviewed and presented with a questionnaire to learn their subjective opinion about those intrinsic design factors of Chinese CAPTCHAs. The satisfactory questionnaire utilized the 5-point Likert-scale (1=strongly disagree, 5=strongly agree) and focused on three aspects: Q1. It is visually comfortable; Q2. It's easy and efficient to recognize and input; Q3. It's appropriate for wide application. Those questionnaire results are displayed in Table 2 and 3. Friedman test results are listed in Table 4, which reveal statistically significant differences between those variables.

**Table 2**. Satisfaction of Chinese CAPTCHAs based on four different fonts: Yahei, Songti, Heiti and Caoshu. The questionnaire used a 5-point Likert-scale (1=strongly disagree, 5=strongly agree).

| | Yahei | | Songti | | Heiti | | Caoshu | |
|---|---|---|---|---|---|---|---|---|
| | AVG | SD | AVG | SD | AVG | SD | AVG | SD |
| **Q1.** It's visually comfortable | 4.43 | 0.86 | 4.10 | 0.99 | 4.40 | 0.81 | 2.60 | 0.93 |
| **Q2.** It's easy and efficient to recognize and input | 4.43 | 0.90 | 4.23 | 0.97 | 4.47 | 0.90 | 2.83 | 1.23 |
| **Q3.** It's appropriate for wide application | 4.10 | 0.82 | 4.10 | 0.92 | 4.40 | 0.86 | 2.43 | 0.86 |

**Table 3**. Satisfaction of Chinese CAPTCHAs based on low frequency characters [LFC], similar form characters [SFC], similar initial consonant characters [SIC] and similar vowel characters [SVC]. The questionnaire used a 5-point Likert-scale (1=strongly disagree, 5=strongly agree).

| | LFC | | CFS | | SIC | | SVC | |
|---|---|---|---|---|---|---|---|---|
| | AVG | SD | AVG | SD | AVG | SD | AVG | SD |
| **Q1.** It's visually comfortable | 3.53 | 1.14 | 3.97 | 1.03 | 4.20 | 1.00 | 4.17 | 0.95 |
| **Q2.** It's easy and efficient to recognize and input | 2.93 | 0.91 | 4.06 | 0.94 | 3.90 | 0.99 | 3.53 | 1.04 |
| **Q3.** It's appropriate for wide application | 2.77 | 0.97 | 3.90 | 0.84 | 3.87 | 0.94 | 3.5 | 1.00 |

**Table 4**. Friedman test results for the variables listed in Table 2 and 3, which indicate statistical differences

| | Yahei, Songti, Heiti, Caoshu | | Yahei, LFC, SFC, SIC, SVC | |
|---|---|---|---|---|
| | Chi-square | Asymp.Sig. | Chi-square | Asymp.Sig. |
| **Q1.** It's visually comfortable | 77.77 | $p<0.001$ | 60.13 | $p<0.001$ |
| **Q2.** It's easy and efficient to recognize and input | 70.93 | $p<0.001$ | 86.19 | $p<0.001$ |
| **Q3.** It's appropriate for wide application | 77.10 | $p<0.001$ | 92.05 | $p<0.001$ |

Among all the four fonts, participants were satisfied with Yahei, Songti and Heiti. The font Caoshu, which is relatively harder to recognize, was not preferred. In addition, CAPTCHAs that contained less-frequently or similar form characters also turned out to be less favored. Although participants reported that characters with similar pronunciations are visually comfortable, they believed that CATPCHAs based on those characters were less efficient. This is probably because that it takes extra efforts to distinguish the similar pronunciations during the typing of those characters

### 5.3. Procedure for the localization of text CAPTCHAs

The localization study of Chinese CAPTCHAs presented here is expected to be capable of generalizing to many other languages, such as Arabic, Japanese, Korean, Indian, Russian, etc. Fig. 9 shows the general procedure we proposed for the localization of text CAPTCHAs, which consists of three consecutive steps: (i) Comparing the usability of CPATCHAs based on English and local language; (ii) Evaluating the design factors that may affect the usability of localized CAPTCHAs; (iii) Refining CAPTCHA designs according to security analysis and medium scale user test.

The goal of the first step is to determine if CAPTCHAs based on local language is comparable or better than English ones. During this stage, the first thing is to analyze the cognition processes of solving CAPTCHAs that employ local language and English. According to those results, a design matrix is generated to help prepare English and local CAPTCHAs within a similar frame for the comparison of usability. For example, selection of fonts that support both languages and are equally popular for local users; determination of the candidate texts such as random characters or frequently-used words, etc. Thereafter, native speakers of local language who are also familiar with English are recruited to participant in the usability test. Because those bilingual participants are familiar with both the English and local characters presented on a CAPTCHA, the intrinsic usability difference between those two languages is evaluated within a similar frame in which the user difference is minimized. For native speakers who are unfamiliar with English, their performance on English CAPTCHA is expected to be lower than those who know English, as is the case for Urdu, the local language in Indian[13]. As a consequence, if the usability test indicates that bilingual participants perform equally or better with local language CAPTCHAs, then it is worthy to localize those CAPTCHAs based on that local language. Otherwise, it is suggested to use English CAPTCHAs.

If CAPTCHAs based on local language provide equal or better usability than English ones, then it is time to move to the next step: evaluating the design factors that may affect the usability of a localized CAPTCHA. Those design factors are classified into two categories: intrinsic factors and general factors. The intrinsic factors are defined as those uniquely related with a particular language, for example, similarity of characters in form or pronunciation, typical font families, etc. All other design factors are classified as general factors, which can be applied to all languages, such as background noise, distortion, overlapping, rotation angle, etc. This study mainly evaluated the intrinsic factors, because the results of Experiment I preliminarily showed that Chinese CAPTCHAs are not significantly affected by general factors like background noise, distortion etc. However, for the full process of CAPTCHA localization, it is recommended to evaluate both the intrinsic and general factors. During the last setup, security evaluation and medium-scale user test are suggested to further polish the design of localized CAPTCHAs, followed by the actual deployment.

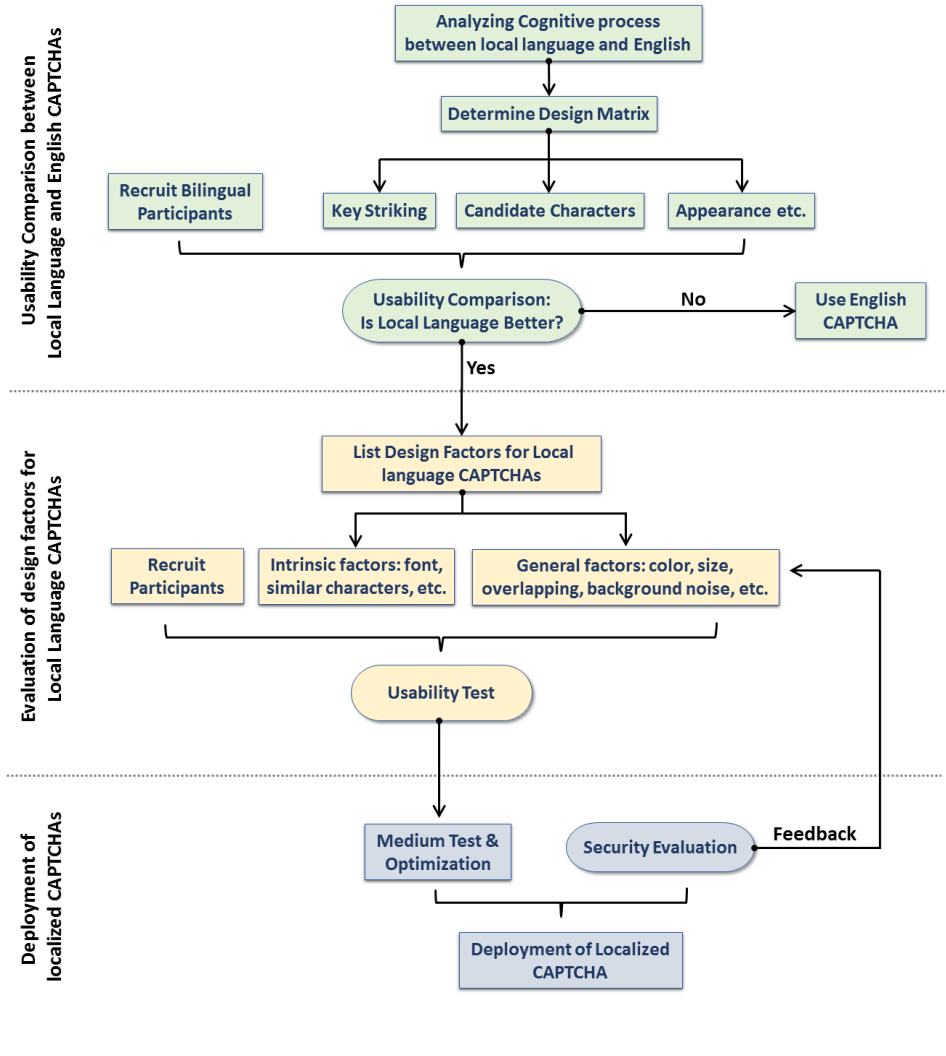

**Fig. 9**. Procedure for the localization of text CAPTCHAs

## VI. Conclusion

In this study, we analyzed the cognition processes of solving English and Chinese CAPTCHAs, compared the usability of CAPTCHAs based on the two languages, evaluated the intrinsic factors that may affect the usability of Chinese CAPTCHAs, and finally proposed a procedure that is applicable for the design of user-friendly CAPTCHAs based on other languages. Although the cognition process of solving Chinese CAPTCHAs involves more efforts due to its less straightforward input method than that of English, the usability comparison experiments indicated that CAPTCHAs based on the two languages are almost equally usable within the same design factors such as font size and family, distortion, background noise level and typing workload. This is mainly attributed to the fact that native speakers are skillful in typing Chinese Characters. Further analyses on intrinsic design factors of Chinese CAPTCHAs revealed that, the usability of those CAPTCHAs is less sensitive to the daily-used fonts such as Yahei, Songti or Heiti. Meanwhile, although the characters that are similar in form didn't affect the usability, characters that are similar in pronunciation may bring confusions to the solving of such CAPTCHAs because it affects the input of those characters. Basing on the comprehensive evaluation on the localization of CAPTCHAs employing Chinese characters, a generalized procedure for the localization practice of other languages was proposed, which includes three steps: usability comparison between alphanumeric and local-language CAPTCHAs, evaluating design factors that may affect the usability of local language CPATCHAs, medium scale deployment for feedback and final deployment of localized CAPTCHAs. This study may shine a light for designing user-friendly CAPTCHAs that employ local languages.

**Limitations of this study and future work**

The limitation of this study falls into the following aspects: The current study mainly focused on the usability while the security of localized CAPTCHAs is not thoroughly discussed. We've utilized commercial optical character recognize (OCR) software to preliminarily test all the CAPTCHAs mentioned in this study and found that those based on Chinese characters were securer than alphanumeric ones. However, in real case, the hack of CAPTCHAs usually includes two consecutive processes, segmentation and recognition [5, 23], which are dedicated to the recognition of distorted characters with background noises [24, 25]. A comprehensive study on the security aspect of Chinese CAPTCHAs with different design factors would be an interesting topic in the future. Another limitation is the scale of the usability test. However, with the rapid development of Amazon Mechanical Turk [26], it is possible to deploy Chinese CAPTCHA designs for large scale evaluations, which is expected to provide more specific results, such as how users of different experience with computers would response to those CAPTCHAs; how people of different ages may perform, etc. Moreover, design factors that are not thoroughly discussed in this study, distortion or background noise for example, could be investigated in detail with such large scale test. Also, the CAPTCHAs studied here is not deployed in real-life situations, such as registrations for an online account or login of Email systems. On possible solution is to incorporate such tests into a real life situation. For example, previous studies have shown the advantages of integrating usability test of CAPTCHAs into real-life system that is design for online courses [27].

**Acknowledgements**. Junnan Yu gratefully thank Dr. Runze Li for helpful discussion. This work was supported by Shanghai Pujiang Program under Grant No. 13PJC072, Shanghai Philosophy and Social Science Program under Grant No. 2012BCK001, and Shanghai Jiao Tong University Interdisciplinary among Humanity, Social Science and Natural Science Fund under Grant No. 13JCY02.